\documentclass[12pt,psfig]{article}
\usepackage{amssymb}
\usepackage{amsmath,amsthm}
\usepackage{amsfonts}
\usepackage{graphicx}
\usepackage{graphics}
\usepackage{indentfirst}
\numberwithin{equation}{section}

\usepackage{hyperref}

\begin{document}
\begin{center}\Large\textbf{LREE of a Dynamical Unstable
D$p$-brane}
\end{center}
\vspace{0.75cm}
\begin{center}{\large Shirin Teymourtashlou and \large Davoud
Kamani}\end{center}
\begin{center}
\textsl{\small{Department of Physics, Amirkabir University of
Technology (Tehran Polytechnic) \\
P.O.Box: 15875-4413, Tehran, Iran \\
e-mails: sh.teymourtash@aut.ac.ir , kamani@aut.ac.ir \\}}
\end{center}
\vspace{0.5cm}

\begin{abstract}

We derive the left-right entanglement entropy 
(LREE) for a bosonic D$p$-brane. This brane has  
tangential dynamics and has been dressed 
by a $U(1)$ gauge potential, the Kalb-Ramond field
and a tachyon field. For this 
purpose, the R\'{e}nyi entropy will be computed and 
then, by taking a special limit of it, the LREE
will be obtained. Besides, the behavior of the LREE 
under the tachyon condensation process
will be evaluated. In addition, after the transition
of the system, i.e. the collapse of the unstable brane,
the second law of the thermodynamics on the 
LREE will be checked.
We find that preserving the second law imposes 
some conditions on the parameters of the setup.

\end{abstract}



\newpage
\section{Introduction}

The entanglement entropy, as an appropriate measure 
for entanglement, has been widely  
studied in the different contexts. For 
instance, this quantity is employed in 
many-body quantum systems to study the quantum 
phases and phase transitions \cite{1}, \cite{2}. 
Besides, connections between the 
entanglement entropy and black hole entropy 
were found \cite{3}, \cite{4}. 
More appealing works were done 
in the AdS/CFT area, which reveal some 
connections between the entanglement entropy 
and gravity \cite{5}, \cite{6}.

Since a boundary state encodes the whole properties of 
the associated D-brane it 
is a suitable tool for studying the D-branes, 
their interactions and so on \cite{7}-\cite{15}. 
In this paper, by applying the boundary 
state formalism, our goal is to 
investigate a special property 
of the D-branes, which is called 
left-right entanglement entropy (LREE) 
\cite{16}-\cite{19}.
Thus, we figure out the LREE of a bosonic unstable D$p$-brane 
with a tangential linear motion and rotation 
which has been dressed by the 
Kalb-Ramond field, a $U(1)$ gauge potential 
and a tachyon field of the open string spectrum.
Traditionally, in order to measure the 
entanglement in a bipartite system 
the system is geometrically divided to subsystems
\cite{20}. In our approach, similar to 
the Ref. \cite{16}, the division occurs 
in the Hilbert space. Precisely, the two  
subsystems are the left- and right-moving 
modes of closed strings which appear in the 
expansion of the boundary state
as a favorable bipartite system. 

On the other hand,
according to the literature, e.g. Ref. \cite{21}, 
presence of the open string tachyon on a 
D-brane obviously makes it unstable. Consequently, 
it decays to an unstable lower 
dimensional D-brane through the 
tachyon condensation process. 
The resultant intermediate brane eventually 
collapses to the closed string vacuum 
or decays to a lower dimensional stable brane
\cite{22}-\cite{27}. 
Since the open string tachyon field lives on our brane 
we were motivated to investigate 
effect of the tachyon condensation on the 
LREE corresponding to our unstable brane. Besides, since the 
similarity of the thermal and entanglement entropies 
has been demonstrated \cite{28}-\cite{31},
we were stimulated to examine the second law of thermodynamics 
for changing the LREE via the condensation of 
the tachyon.

In fact, the corresponding LREE of a D-brane
potentially possesses a connection with the 
entropies of the black holes
\cite{3, 4}. Therefore, the LREE
of our setup may find a relation with the
entropies of the rotating-moving charged black holes.

Note that the LREE was originally studied by 
P. Zayas and N. Quiroz for a 
one-dimensional boundary state in a 
2D CFT with the Dirichlet or the Neumann 
boundary condition \cite{16}. Then, 
they extended their work to the case of 
a bare-stationary D$p$-brane \cite{17}. 
We shall apply their approach to 
compute the LREE of our setup. 
 
The paper is organized as follows. 
In Sec. 2, we shall introduce the boundary state, 
associated with a dynamical-dressed unstable 
D$p$-brane. Then the interaction 
amplitude between two parallel D$p$-branes, which is 
necessary for the calculation 
of the LREE, will be introduced. 
In Sec. 3, we compute the LREE for the foregoing  
D$p$-brane. 
In Sec. 4, the effect of the tachyon condensation
on the LREE will be calculated, and some 
thermodynamical interpretations will be presented. 
Section 5 is devoted to the conclusions.

\section{The boundary state and interaction amplitude}

\subsection{The boundary state}

Consider a D$p$-brane with tangential 
dynamics in the presence of the antisymmetric
field $B_{\mu \nu}(X)$, a $U(1)$ gauge 
potential $A_{\alpha}(X)$ and an open 
string tachyon field $T(X)$. In order to introduce 
the corresponding boundary 
state, we start with the action
\begin{eqnarray}
S=&-&\frac{1}{4\pi\alpha'}\int_{\Sigma} {\rm d}^2\sigma
\left(\sqrt{-g} g^{ab}G_{\mu\nu} \partial_{a}X^\mu
\partial_b X^{\nu} +\varepsilon^{ab}
B_{\mu\nu}\partial_a X^{\mu}
\partial_b X^{\nu}\right)
\nonumber\\[10pt]
&+&\frac{1}{2\pi\alpha'}\int_{\partial\Sigma}
{\rm d}\sigma\left(A_{\alpha}
\partial_{\sigma}X^{\alpha}
+\omega_{\alpha\beta}J^{\alpha\beta}_{\tau}
+T(X^\alpha) \right),\label{eq:2.1}
\end{eqnarray}
where we shall use the sets 
$\{x^\alpha|\alpha=0,1,\cdots,p\}$ and 
$\{x^i|i=p+1,\cdots,d-1\}$ to show the 
parallel and perpendicular directions 
to the brane worldvolume, respectively. 
We apply the reliable gauge 
$A_\alpha=-\frac{1}{2}F_{\alpha \beta} X^\beta$ 
with the constant field strength $F_{\alpha\beta}$, 
and adopt the tachyon profile as 
$T=\frac{1}{2} U_{\alpha \beta}X^\alpha X^\beta$ 
with the constant symmetric matrix $U_{\alpha\beta}$. 
The spacetime and worldsheet metrics and the Kalb-Ramond field
are taken to be constant with  
$G_{\mu \nu}=\eta_{\mu\nu}=\rm{diag}(-1,1,...,1)$.
The antisymmetric matrix $\omega_{\alpha\beta}$ 
denotes the spacetime angular velocity of the brane inside 
its worldvolume and 
$J^{\alpha\beta}_\tau=X^\alpha \partial_\tau X^\beta
-X^\beta \partial_\tau X^\alpha$ represents 
the angular momentum density.

Comparing the setup of this paper with that 
of our previous work \cite{32}, we can say
the branes of both papers have been dressed 
by the same Kalb-Ramond field and the same 
internal $U(1)$ gauge potential. Besides, both branes
have the same tangential dynamics. The only 
difference between them is the instability
of the present brane, which is
induced by the tachyon field. We shall observe 
that the associated LREE of the unstable brane
is very different from the LREE of the stable 
brane.

Variation of the action with respect 
to $X^\mu$ gives the equation of motion 
and following boundary state equations
\begin{eqnarray}
&~&\left( \Delta_{\alpha\beta} \partial_\tau X^\beta
+\mathcal{F}_{\alpha\beta} 
\partial_{\sigma} X^\beta+B_{\alpha i} 
\partial_\sigma X^i+U_{\alpha\beta}
X^\beta \right)_{\tau=0} |B_x\rangle=0,
\nonumber\\[10pt]
&~&\left(X^i-y^i\right)_{\tau=0}|B_x\rangle=0,
\end{eqnarray}
where $\Delta_{\alpha\beta}=\eta_{\alpha\beta}
+4\omega_{\alpha\beta}$ 
and $\mathcal{F}_{\alpha\beta}=B_{\alpha\beta}
-F_{\alpha\beta}$.
The transverse vector $y^i$ determines
the brane position. Using 
the mode expansion of $X^\mu$ for 
the propagating closed string and the decomposition 
$|B_x\rangle = |B_{(\rm osc)}\rangle \otimes |B_{(0)}\rangle$ 
one can rewrite these
equations in terms of the string oscillators
\begin{eqnarray}
&~&\left[ \left(\Delta_{\alpha\beta}
-\mathcal{F}_{\alpha\beta}+\frac{i}{2m} 
U_{\alpha\beta}\right) \alpha^\beta_m
+ \left(\Delta_{\alpha\beta}+
\mathcal{F}_{\alpha\beta}-\frac{i}{2m} 
U_{\alpha\beta}\right) 
\tilde{\alpha}^\beta_{-m}\right]|B_{(\rm osc)}\rangle=0,
\nonumber\\[10pt]
&~&\left(2\alpha'\Delta_{\alpha\beta}\;p^\beta
+U_{\alpha\beta}\;x^\beta\right)
|B_{(0)}\rangle=0, \label{eq:2.3}
\end{eqnarray}
for the tangential directions and
\begin{eqnarray}
&~& (\alpha^i_m-\tilde{\alpha}^i_{-m})
|B_{(\rm osc)}\rangle=0,
\nonumber\\
&~& (x^i-y^i)|B_{(0)}\rangle=0,
\end{eqnarray}
for the perpendicular directions.  

Applying the quantum mechanical methods, 
particularly the coherent state formalism, 
the zero-mode and oscillating parts of 
the boundary state find the features
\begin{eqnarray}
|B_{(0)}\rangle&=&\frac{T_p}{2\sqrt{\det(U/4\pi \alpha')}} 
\int_{-\infty}^{\infty} \prod^p_{\alpha=0}
\exp\bigg[ i\alpha' \sum_{\beta \neq \alpha}
(U^{-1}\Delta+\Delta^{\rm T}\; 
U^{-1})_{\alpha\beta}p^\alpha p^\beta
\nonumber\\[10pt]
&+&\frac{i \alpha'}{2}(U^{-1}\Delta
+\Delta^{\rm T}\; U^{-1})_{\alpha\alpha}
(p^\alpha)^2 \bigg] |p^\alpha \rangle {\rm d}p^\alpha 
\nonumber\\[10pt]
&\times& \prod_{i=p+1}^{d-1} 
\left[ \delta(x^i-y^i) |p^i=0\rangle \right],\label{eq:2.5}
\\[10pt]
|B_{(\rm osc)}\rangle &=& 
\prod_{n=1}^{\infty} [-\det M_{(n)}]^{-1} 
\exp\left[ -\sum_{m=1}^{\infty} 
\left( \frac{1}{m} \alpha^{\mu}_{-m} 
S_{(m)\mu\nu} \tilde{\alpha}^\nu _{-m} \right) \right] 
|0\rangle_\alpha |0\rangle_{\tilde \alpha},\label{eq:2.6}
\end{eqnarray}
where $T_p$ is the tension of the D$p$-brane. 
The matrix $S_{(m)\mu\nu}$ is defined by
$S_{(m)\mu\nu}=(Q_{(m)\alpha\beta}\equiv(M_{(m)}^{-1}
N_{(m)})_{\alpha\beta},-\delta_{ij})$, 
in which
\begin{eqnarray}
M_{(m)\alpha\beta}=\Delta_{\alpha\beta}
-\mathcal{F}_{\alpha\beta}
+\frac{i}{2m}U_{\alpha\beta},
\nonumber\\
N_{(m)\alpha\beta}=\Delta_{\alpha\beta}
+\mathcal{F}_{\alpha\beta}
-\frac{i}{2m}U_{\alpha\beta}.
\end{eqnarray}
In fact, the parameters of the setup, which 
were appeared in the boundary state 
$|B_x\rangle$, are not independent.
That is, the first equation of Eqs. (\ref{eq:2.3}) 
enables us to express the left- and right-moving 
oscillators in terms of each other. 
If we choose the set 
$\{\alpha^\alpha_m ,{\tilde \alpha}^\alpha_{-m}|m 
\in \mathbb{N}\}$ 
the boundary state manifestly possesses the matrix  
$Q_{(m)\alpha\beta}$, and if we select the set
$\{{\tilde \alpha}^\alpha_m ,\alpha^\alpha_{-m}|m 
\in \mathbb{N}\}$ 
it will contain the new matrix  
$\left( \left[Q_{(-m)}^{-1}\right]^{\dagger} 
\right)_{\alpha\beta}$. Equality 
of these matrices leads to the condition 
$Q_{(m)} Q^{\dagger}_{(-m)}=\textbf1$, which gives
\begin{eqnarray}
&~& \Delta\; U=U\; \Delta^{\rm T},
\nonumber\\
&~&\Delta\; \mathcal{F}=\mathcal{F}\; 
\Delta^{\rm T}. \label{eq:2.8}
\end{eqnarray}

The normalization prefactors of 
Eqs. (\ref{eq:2.5}) and (\ref{eq:2.6}) come 
from the disk partition function. For example, 
look at the Eq. (\ref{eq:2.6}).
For the constant Kalb-Ramond field the second 
term of the action (\ref{eq:2.1}) reduces to a boundary 
term. Therefore, we receive a total boundary 
action which includes the matrices 
${\cal{F}}_{\alpha \beta}$, $\omega_{\alpha \beta}$
and $U_{\alpha \beta}$. By computing 
the partition function via the 
boundary action the quantities 
$\{\det M_{(m)}| m = 1, 2, 3, \cdots\}$
appear in the normalization factor.
Similar normalization factors can be found, e.g., 
in Refs. \cite{14, 15}.

Beside the matter part of the boundary state, there  
is a contribution by the conformal ghosts too,
\begin{equation}
|B_{\rm gh}\rangle=\exp \left[ \sum_{n=1}^\infty 
(c_{-n}\tilde b_{-n}-b_{-n} \tilde c_{-n})\right] 
\frac{c_0+\tilde c_0}{2}\; |q=1\rangle\; |\tilde q=1\rangle.
\end{equation}
Thus, the total bosonic boundary state takes the form
\begin{equation}
|B\rangle=|B_{(\rm osc)}\rangle \otimes |B_{(0)}\rangle 
\otimes |B_{\rm gh}\rangle. \label{eq:2.10}
\end{equation}

\subsection{The interaction amplitude}

The calculation of the left-right entanglement 
entropy needs to extract the partition function.
Thus, we introduce the interaction amplitude of 
two dynamical D$p$-branes which are dressed 
by the foregoing fields. For this purpose, 
we consider the tree-level 
diagram of a closed string which propagates 
between two such D$p$-branes.
The amplitude can be computed by the overlap 
of the total boundary states $|B_{1,2}\rangle$
via the propagator $D$ of the exchanged closed string 
\begin{eqnarray}
\mathcal{A}&=&\langle B_1|D|B_2\rangle,
\nonumber\\
D&=&2 \alpha' \int_0^\infty {\rm d}t\;e^{-tH},
\end{eqnarray}
where $H$ is the total Hamiltonian of the 
propagating closed string, including the ghost part. 
Hence, the amplitude is given by 
\begin{eqnarray}
\mathcal{A}&=&\frac{T_p^2 \alpha' 
V_{p+1}}{16 (2\pi)^{d-p-1}}\;
\frac{\prod_{m=1}^\infty \left[\det 
\left(M_{(m)1}^\dagger M_{(m)2}
\right) \right]^{-1}}{\sqrt{\det(U_1/{4\pi\alpha'}) 
\det(U_2/{4\pi\alpha'})}}\;
\nonumber\\[10pt]
&\times& \int_0^\infty {\rm d}t \bigg\{ e^{(d-2)\pi t/6}
\left(\frac{1}{\sqrt {\alpha' t}}\right)^{d-p-1} 
\exp \left(-\frac {1}{4\alpha' t} 
\sum_{i=p+1}^{d-1}(y_1^i-y_2^i)^2\right)
\nonumber\\[10pt]
&\times&\prod_{m=1}^\infty \bigg( \det 
\left[ \textbf 1-Q_{(m)1}^\dagger Q_{(m)2} 
e^{-4m\pi t} \right]^{-1} 
\left(1-e^{-4m\pi t}\right)^{p-d+3} 
\bigg) \bigg \}, \label{eq:2.12}
\end{eqnarray}
where $V_{p+1}$ represents the worldvolume of
each brane.
The first exponential comes from the 
zero-point energy, the two factors next to it 
originate from zero modes. In addition, 
the determinant part in the last line is the 
contribution of the Neumann oscillators while the factor 
$\prod_{m=1}^\infty \left(1-e^{-4m\pi t}\right)^{p-d+3}$ 
is due to the Dirichlet oscillators and the conformal ghosts.

\section{The corresponding LREE to the 
dynamical-dressed unstable D$p$-brane}

In a composite quantum system, which includes
some subsystems,
entanglement clearly relates the various parts
of the system. The quantum state of each subsystem
is not independent of the states of the
other subsystems. 
Entanglement entropy is an appropriate quantity
for measuring the entanglement among the subsystems.

Consider a bipartite system which comprises  
the subsystems A and B. 
If the pure state of the composite system 
is denoted by $|\psi \rangle$, 
then the density operator of this state is given by 
$\rho=|\psi \rangle\langle \psi |$, which  
satisfies the probability conservation
${\rm Tr} \rho=1$. Therefore, 
the reduced density operator 
for the subsystem A is given by the partial 
trace over the subsystem B, i.e., 
$\rho_{\rm A}={\rm Tr}_{\rm B}\rho$.

For measuring the entanglement, we choose the 
entanglement and R\'{e}nyi entropies. 
The former quantity is given by the 
von Neumann formula 
$S=-{\rm Tr}(\rho_{\rm A} \ln \rho_{\rm A})$ 
\cite{33}, and the latter is defined by 
$S_n=\frac{1}{1-n} \ln {\rm Tr}\rho_{\rm A}^n$, 
where $n\geq 0$ and $n\neq 1$. By taking the limit $n \to 1$, 
the entanglement entropy can 
be derived from the R\'{e}nyi entropy \cite{34}.

\subsection{The density operator of the configuration}

As we know the Hilbert space of string theory 
can be represented by the direct product of two 
subspaces L and R with the left- and right-moving modes 
as its bases. That is, the Hilbert space possesses 
the factorized form 
$\mathcal H=\mathcal H_{\rm L} \otimes \mathcal H_{\rm R}$.
By imposing the Virasoro constraints, which leads to the 
same excitation number on the left and right sectors,
we receive the physical Hilbert space. 
Therefore, these constraints  
very weakly relate the left and right sectors,
and in principle they are still independent. 
Explicitly, the most general state for the 
closed strings has the feature 
$|\psi\rangle=|\psi\rangle_{\rm L} \otimes |\psi\rangle_{\rm R}$, 
where
\begin{eqnarray}
|\psi\rangle_{\rm L}&=&\prod_{k=1}^\infty 
\frac{1}{\sqrt{n_k !}}
\left(\frac{\alpha^{\mu_k}_{-k}}{\sqrt{k}}
\right)^{n_k}|0 \rangle ,
\nonumber\\
|\psi\rangle_{\rm R}&=&\prod_{k=1}^\infty 
\frac{1}{\sqrt{m_k !}}
\left(\frac{\tilde{\alpha}^{\mu_k}_{-k}}{\sqrt{k}}
\right)^{m_k}|0\rangle,
\nonumber\
\end{eqnarray}
such that $\sum_{k=1}^\infty kn_k=\sum_{k=1}^\infty km_k$.
We see that the sets 
$\{n_k|k=1,2,3, \cdots\}$ and $\{m_k|k=1,2,3, \cdots\}$,
up to the mentioned constraint, are 
essentially independent. This implies that the 
physical Hilbert space is still a product of 
the left and right sectors. 

Besides, a boundary state satisfies the physical constraint 
$(\mathcal L_n-\tilde {\mathcal L}_{-n})|B\rangle=0$ 
for any $n\in \mathbb{Z}$, where $\mathcal L_n$ and 
$\tilde{\mathcal L}_{-n}$ are the Virasoro operators. 
However, the boundary state can be decomposed  
to the left- and right-moving modes 
via Schmidt procedure \cite{35, 36}. Thus, we can take 
the total boundary state (\ref{eq:2.10}) as our composite 
system. Since  the matrix $S_{(m)\mu\nu}$ in Eq. (\ref{eq:2.6}) 
has nontrivial elements, our system is an entangled composite 
system with the left-right entanglement.

For a given boundary state $|B\rangle$ 
if we define the density matrix as 
$\rho=|B\rangle \langle B|$, because of the 
divergence of the inner product $\langle B|B\rangle$,
we don't receive the probability conservation 
$\rm Tr \rho=1$. Hence, we 
have to introduce a finite correlation length 
$\epsilon$ to acquire  
the condition $\rm Tr \rho=1$, \cite{29, 37}.
Thus, the density matrix is redefined by
\begin{equation}
\rho= \frac{e^{-\epsilon H}|B\rangle 
\langle B|e^{-\epsilon H}}{Z(2\epsilon)},
\end{equation}
where the denominator is fixed by the 
probability condition as
\begin{eqnarray}
Z(2\epsilon)&=&\; \langle B|e^{-2\epsilon H}|B\rangle
\nonumber\\[10pt]
&=&\frac{T_p^2\; V_{p+1}}{8(2\pi)^{d-p-1}}\;
\frac{\prod_{m=1}^{\infty}
|\det M_{(m)}|^{-2}}{\det(U/{8\pi})} 
e^{(d-2)\pi \epsilon/3}
\left( \frac{1}{2\sqrt{\epsilon}}\right)^{d-p-1}
\nonumber\\[10pt]
&\times&\prod_{m=1}^\infty 
\bigg(\det \left[ \textbf{1}-Q_{(m)}^\dagger 
Q_{(m)} e^{-8m\pi \epsilon} \right]^{-1}
(1-e^{-8m\pi \epsilon})^{p-d+3} \bigg) .\label{eq:3.2}
\end{eqnarray}
Eq. (\ref{eq:2.12}) implies that $Z(2\epsilon)$ 
is the tree-level amplitude 
of a closed string which propagates for 
the time $t=2\epsilon$ between the 
two identical D$p$-branes.
Since the D$p$-branes are identical and have been 
located in the same position 
the indices 1 and 2 were omitted and 
the $y$-dependence was also removed.
Note that we have taken $\alpha'=2$.

\subsection{The LREE of the setup}

The reduced density matrix for the subsystem L, 
i.e. $\rho_{\rm L}$, is obtained by taking 
the trace over the right-moving oscillators. 
In order to find the R\'{e}nyi entropy, we 
apply the replica trick which simplifies 
the computation of ${\rm Tr}\rho_{\rm L}^n$ 
with real $n$, 
\begin{equation}
{\rm Tr}\rho_{\rm L}^n\; \sim 
\frac{Z(2n\epsilon)}{Z^n (2 \epsilon )}
\equiv \frac{Z_{n}({\rm L})}{Z^n}~,
\end{equation}
where $Z_n({\rm L})$ is the replicated form 
of the partition function. Note that we shall use the 
partition function (\ref{eq:3.2}) to calculate the 
above relation, and then we express it in 
terms of the Dedekind eta-function
\begin{equation}
\eta(q)=q^{1/12}\prod_{m=1}^\infty(1-q^{2m}),
\end{equation}
where $q=e^{-4\pi \epsilon}$. In the 
limit $\epsilon \to 0$ the variable $q$ approaches to 
one. To avoid this value of $q$
we apply the open/closed worldsheet 
duality via the transformation 
$4\epsilon \to 1/4\epsilon$ 
to go to the open string channel. Consequently, 
as $\epsilon$ goes to zero the 
new variable $\tilde{q}=e^{-\pi/4\epsilon}$ 
vanishes and we can utilize 
the expansion of the Dedekind $\eta$-function for 
small argument. Therefore, we receive  
the following relation
\begin{eqnarray}
\frac{Z_n}{Z^n}&\approx&\;K_p^{1-n} \; 
\left(\left(2\sqrt {\epsilon}\right)^{1-n}\; 
\sqrt{n}\right)^{d-p-1}\; \exp \left[\frac{(d-2)\pi}
{48\epsilon}\left(\frac{1}{n}-n\right)\right]
\nonumber\\[10pt]
&\times& \prod_{m=1}^\infty \bigg\{1+(d-p-3)
\bigg[ -n\;e^{-m\pi/2\epsilon}
+\frac{n^2}{2} \left(d-p-3-\frac{1}{n}\right)
e^{-m\pi/\epsilon}
\nonumber\\[10pt]
&+&e^{-{m\pi}/{2\epsilon n}}-n(d-p-3)\;
e^{-(1+1/n){m\pi}/{2\epsilon}}
+\frac{d-p-2}{2}\; e^{-{m\pi}/{\epsilon n}}
\bigg{]}\bigg{\}}
\nonumber\\[10pt]
&\times& \prod_{m=1}^{\infty}\bigg{\{} 1
-n\; {\rm Tr} \left(Q_{(m)}^{\dagger} Q_{(m)}\right)\;
e^{-{m\pi}/{2\epsilon}} -n \;\left[{\rm Tr} 
\left(Q_{(m)}^{\dagger} Q_{(m)}\right)\right]^2 \;
e^{-m\pi( {1}/{\epsilon}+{1}/{n\epsilon})}
\nonumber\\[10pt]
&+& {\rm Tr} \left(Q_{(m)}^{\dagger} Q_{(m)}\right) \;
e^{-{m\pi}/{2n\epsilon}}
+\frac{n}{2} \bigg{[}
-{\rm Tr} \left(Q_{(m)}^{\dagger} Q_{(m)}\right)^{2}
+ n \left[{\rm Tr} \left(Q_{(m)}^{\dagger} 
Q_{(m)}\right)\right]^2 \bigg{]}e^{-{m\pi}/{\epsilon}}
\nonumber\\[10pt]
&+& \frac{1}{2}\bigg{[} {\rm Tr} 
\left(Q_{(m)}^{\dagger} Q_{(m)}\right)^2 \; +
\left[{\rm Tr} \left(Q_{(m)}^{\dagger} 
Q_{(m)}\right)\right]^2 
\bigg{]}e^{-{m\pi}/{n\epsilon}}\bigg{\}}~,\label{eq:3.5}
\end{eqnarray}
up to the order 
$\mathcal {O}(\exp(-3\pi /2\epsilon))$.
The factor $K_p$ was entered into 
Eq. (\ref{eq:3.5}) via Eq. (\ref{eq:3.2}),
and possesses the following definition 
\begin{equation}
K_p\;=\;\frac{T_p^2\; V_{p+1}}{8\;(2\pi)^{d-p-1}}\;
\frac{\prod_{m=1}^{\infty}|\det M_{(m)}|^{-2}}{\det(U/{8\pi})}.\label{eq:3.6}
\end{equation}
Now, by taking the limit $n \to 1$ of the R\'{e}nyi entropy 
we acquire the entanglement entropy
\begin{eqnarray}
S_p&\approx& \ln K_p +\frac{d-p-1}{2} 
\; (2\ln 2+\ln \epsilon -1)
+ \frac{(d-2)\pi}{24\epsilon} 
\nonumber\\[10pt]
&+&\sum_{m=1}^\infty \bigg\{
\left[ {\rm Tr}\left(Q_{(m)}^{\dagger} 
Q_{(m)}\right)+d-p-3\right]\;
\left(1-\frac{m\pi}{2\epsilon}\right) 
e^{-m\pi/2\epsilon}
\nonumber\\[10pt]
&+&\left[ {\rm Tr}\left(Q_{(m)}^{\dagger} 
Q_{(m)}\right)^2+d-p-3 \right]
\left(\frac{1}{2}-\frac{m\pi}{2\epsilon}\right)
\; e^{-m\pi/\epsilon} \bigg\}. \label{eq:3.7}
\end{eqnarray}
The first term is concerned to the 
boundary entropy of the brane.
The second term comes from the 
zero-modes, and the next terms are 
contributions of the oscillators 
and conformal ghosts. 
The brane dynamics and the background fields  
were accumulated in the first term and 
series. One of the prominent effects 
of the tachyonic field is the appearance 
of all mode numbers $\{m \in \mathbb{N}\}$
in the LREE. 

By quenching the tachyon field 
the mode dependence of the matrices $Q_{(m)}$ 
and $M_{(m)}$
disappears. In this case by employing the formulas 
$\sum_{m=1}^\infty x^m=\frac{x}{1-x}$ 
and $\sum_{m=1}^\infty mx^m=\frac{x}{(1-x)^2}$
for small $x$ where $x=e^{-\pi/2\epsilon}$ and  
$x=e^{-\pi/\epsilon}$ for the second 
and third lines of Eq. (\ref{eq:3.7}), respectively,
the LREE, up to the term $\ln K_p$,
exactly reduces to Eq. (3.7) of Ref. \cite{32}. 
Now return to the term $\ln K_p$.
In fact, by assuming $\det U \neq 0$
the inverse of the matrix $U$ was appeared in Eq. (\ref{eq:2.5}).
Thus, in the subsequent formulas, e.g. Eq. (\ref{eq:3.7}),
we are not allowed to put the tachyon matrix away.
However, by applying the regularization
scheme $\prod^\infty_{m=1} a \to 1/\sqrt{a}$
the infinite product in Eq. (\ref{eq:3.6}) for $U \to 0$
reduces to $|\det M|$ of Eq. (3.7) of 
the Ref. \cite{32}. Hence, for the case 
$U \to 0$ the term $\ln K_p$ also is
consistent with its counterpart in Ref. \cite{32}.

By turning off all background fields and 
the brane dynamics we acquire the LREE of a 
bare-static D$p$-brane, see Sec. 3.3 of 
Ref. \cite{32}. In this special setup, 
for the configuration $p = 1$
and $d = 3$, the leading terms are exactly 
compatible with Ref. \cite{16}.
In fact, the case $p = 1$ and $d = 3$
is the simplest setup which was originally 
considered for computing the LREE in Ref. \cite{16}.
It represents a warm-up calculation, and is
corresponding to a one-dimensional 
boundary state in a 3-dimensional target spacetime.

\subsection{Connection between the LREE
and thermodynamic entropy}

By defining a temperature, proportional to the inverse of 
the correlation length $\epsilon$, we can specify the thermal 
interpretation of our system. The partition function 
of the dressed-dynamical brane was 
defined by Eq. (\ref{eq:3.2}).
In the limit of $\beta=2\epsilon \to 0$, which 
exhibits the high temperature of the thermal 
system, the thermodynamic entropy of the system is given by
\begin{eqnarray}
S_{\rm {th}}&=&\beta^2\; \frac{\partial}{\partial \beta} 
\left(-\frac{1}{\beta} \ln Z \right)
\nonumber\\[10pt]
&\approx& \ln K_p +\frac{d-p-1}{2} 
\; \left(2\ln 2+\ln \frac{\beta}{2} -1 \right)
+ \frac{(d-2)\pi}{12\beta} 
\nonumber\\[10pt]
&+&\sum_{m=1}^\infty \bigg\{
\left[ {\rm Tr}\left(Q_{(m)}^{\dagger} 
Q_{(m)}\right)+d-p-3\right]\;
\left(1-\frac{m\pi}{\beta}\right) 
e^{-m\pi/\beta}
\nonumber\\[10pt]
&+&\left[ {\rm Tr}\left(Q_{(m)}^{\dagger} 
Q_{(m)}\right)^2+d-p-3 \right]
\left(\frac{1}{2}-\frac{m\pi}{\beta}\right)
\; e^{-2m\pi/\beta} \bigg\},\label{eq:3.8}
\end{eqnarray}
up to the order $\mathcal O(\exp(-{3\pi}/{\beta}))$. 
This demonstrates that the thermal entropy of our 
system exactly matches with its LREE counterpart,
i.e. (\ref{eq:3.7}). There are some other 
papers which also reveal such connections, e.g. 
see the Refs. \cite{28}-\cite{31}.

Due to the presence of the tachyon field 
this thermal entropy has a more general form than 
that of the Ref. \cite{32}.			
We observed that the LREE of our setup, for the  
special case $U \to 0$, reduced to the LREE of 
the stable dressed-dynamical brane \cite{32}.
With the same logic and mathematical tools,
the thermal entropy (\ref{eq:3.8}) for a vanishing tachyon field 
also is consistent with that of the stable brane \cite{32}.

\section{Effect of the tachyon condensation on the LREE}

According to the literature, e.g. Ref. \cite{21}, 
adding an open string tachyonic mode 
to a single D-brane or to a group of D-branes 
makes them unstable. That is, 
in the tachyon condensation process 
the brane dimension is decreased 
\cite{22}, so that eventually there 
will remain the closed string vacuum or
an intermediate stable brane.
In the tachyon condensation phenomenon at 
least one of the elements of the  
tachyon matrix $U_{\alpha\beta}$ becomes infinite. 
For simplicity we impose the condensation 
of the tachyon field only in the $x^p$-direction,
i.e., we apply the limit $U_{pp} \to \infty$.

For simplification, at first we obtain  
the LREE in the limit of large tachyon 
field, i.e., we apply $U \gg 2m(\Delta-\mathcal F)$ 
with the consideration of the conditions (\ref{eq:2.8}). 
Thus, we receive
\begin{eqnarray}
{\tilde S}_p &\approx&\ln K_p 
+\frac{d-p-1}{2} \; (2\ln 2+\ln \epsilon -1)
+ \frac{(d-2)\pi}{24\epsilon} 
\nonumber\\[10pt]
&+&\sum_{m=1}^\infty \bigg\{
\left[d-2-512\;m^2 \;{\rm Tr}
\left(\omega^2 U^{-2}\right)\right]\;
\left(1-\frac{m\pi}{2\epsilon}\right)
e^{-{m\pi}/{2\epsilon}}
\nonumber\\[10pt]
&+&\left[d-2-1024\;m^2\;{\rm Tr}
\left(\omega^2 U^{-2}\right)\right]\;
\left(\frac{1}{2}-\frac{m\pi}{2\epsilon}\right)
e^{-{m\pi}/{\epsilon}}\bigg\}, \label{eq:4.1}
\end{eqnarray}
up to the order $\mathcal {O}(U^{-3})$.
We observe that the large tachyon field 
conveniently quenches the total field strength, unless 
in the first term. 

Now we condensate the tachyon in the $x^p$-direction. 
The first term of Eq. (\ref{eq:4.1}) takes the limit 
\begin{eqnarray}
\lim_{U_{pp} \to \infty} \ln K_p 
&=&\ln (2\pi L_p)+\ln K_{p-1}
\nonumber\\
&\equiv &\gamma + \ln K_{p-1},
\end{eqnarray}
where $L_p$ is the length of the $x^p$-direction of 
the brane. For obtaining this 
we utilized the reliable relation $T_p=T_{p-1}
(2\pi \sqrt{\alpha'})$ and the regularization scheme
$\prod_{n=1}^\infty n \to \sqrt{2\pi}$. 
The second term of Eq. (\ref{eq:4.1}) can be rewritten as
\begin{eqnarray}
\frac{d-p-1}{2} \;(2\ln 2+\ln \epsilon-1)
&=& \frac{d-(p-1)-1}{2}\;(2\ln 2
+\ln \epsilon-1)+\Gamma .
\end{eqnarray}
Moreover, the limit of the factor ${\rm Tr}
(\omega^2 U^{-2})$ is given by 
${\rm Tr}(\; \omega^2 U^{-2})'$ 
where the prime represents a $p \times p$ matrix. 
Adding all these together we acquire 
\begin{eqnarray}
\lim_{U_{pp} \to \infty}{\tilde S}_p
&=&{\tilde S}_{p-1}+\lambda ,
\nonumber\\
\lambda &\equiv & \ln (\pi L_p)-\frac{1}{2}(\ln \epsilon-1).
\end{eqnarray}
After the tachyon condensation the D$p$-brane 
reduces to a D$(p-1)$-brane. The  
corresponding entanglement entropy of the resultant 
brane is specified by ${\tilde S}_{p-1}$. 
The extra entropy $\lambda$ may be associated 
to the created closed strings during the 
collapse of the initial brane. 
The structure of the entropy $\lambda$ elaborates  
that for the larger (smaller) value of the brane length $L_p$  
there is a greater (smaller) number 
of the released closed strings and 
consequently we obtain the larger (smaller) entropy $\lambda$. 

\subsection{The LREE and second law of thermodynamics}

Due to the similar behavior of 
the thermal entropy and the LREE 
\cite{28}-\cite{31}, we were  
motivated to examine the second law of 
thermodynamics for our LREE in the tachyon 
condensation process. During this 
evolution the system goes 
from an initial state, which is the initial D$p$-brane, 
to a final state, i.e., the subsequent D$(p-1)$-brane 
and the released closed strings. 
Hence, the LREE changes 
from the initial entropy $S_i$ to the final entropy $S_f$,
\begin{eqnarray}
S_i&=&{\tilde S}_p ,
\nonumber\\
S_f&=&\lim_{U_{pp} \to \infty} 
{\tilde S}_p={\tilde S}_{p-1}+\lambda .
\end{eqnarray}
In order to check the second law, i.e. 
$S_f > S_i$, we should verify the validity of 
the inequality  
${\tilde S}_{p-1}+\lambda -{\tilde S}_p > 0$,
in which 
\begin{eqnarray}
{\tilde S}_{p-1}+\lambda -{\tilde S}_p&=&\ln \pi 
-\ln\left(\frac{\det U'}{\det U}\right)
-\sum^\infty_{m=1}\bigg \{ \ln \left(\frac{\det 
{M'}_{(m)}}{\det M_{(m)}}\right)^2
\nonumber\\[10pt]
&-&512\; m^2 \; \left[{\rm Tr}\left(\omega^2 U^{-2}\right)
-{\rm Tr}\left(\omega^2 U^{-2}\right)'\right]
\nonumber\\[10pt]
&\times& \left[\left(1-\frac{m \pi}
{2 \epsilon}\right)\; e^{-{m\pi}/{2\epsilon}}
+\left(1-\frac{m \pi}{\epsilon}\right)\; 
e^{-{m\pi}/{\epsilon}}\right]\bigg{\}}, \label{eq:4.6}
\end{eqnarray}
where the primes show the $p \times p $
matrices. Positivity of this quantity for 
$\epsilon \to 0$ imposes the following condition on the 
parameters of the setup
\begin{eqnarray}
\det \left(U \prod^\infty_{m=1}M^2_{(m)} \right) > 
\frac{1}{\pi}\det \left(U' 
\prod^\infty_{m=1}{M'}^2_{(m)} \right). \label{eq:4.7}
\end{eqnarray} 
In fact, this is a minimal condition for preserving 
the second law of thermodynamics. In other words,
the whole quantity in the left-hand-side of Eq. (\ref{eq:4.6})
for any arbitrary and small value 
of ``$\epsilon$'' should be positive.

As a simple example, look at a stretched D1-brane
along the $x^1$-direction. Because of the Lorentz 
symmetry this object can not possess tangential dynamics, 
thus, we have $\omega=0$. There exists a total electric field 
$\mathcal E$ along this D-string. 
After the tachyon condensation along the direction $x^1$, 
there will remain a D0-brane 
and a collection of the released closed strings. 
Accordingly, the Eq. (\ref{eq:4.7}) 
eventuates to the following restriction
\begin{equation}
U_{11}\; > \; \frac{2}{\pi} \; 
\ln \left(\frac{1+\sqrt{1+4/\pi}}{2}\right). 
\end{equation}
Note that in the previous section
we have explicitly chosen very large matrix elements 
for the tachyon  matrix. Therefore, 
the above condition has been already satisfied.

\section{Conclusion}

We computed the left-right entanglement 
entropy of a non-stationary 
D$p$-brane in the presence of an internal 
$U(1)$ gauge potential, the Kalb-Ramond 
field and a tachyon field. For obtaining this, we 
employed the boundary state formalism
and the amplitude of interaction between two
parallel and identical dynamical-dressed D$p$-branes.
Presence of the various parameters in the 
configuration dedicated 
a generalized form to the LREE. Thus, the   
LREE can be accurately adjusted to any desirable value
via these parameters.

By applying the partition function we conveniently 
computed the thermodynamical
entropy, associated with the dressed-dynamical 
unstable D$p$-brane. We compared this entropy 
with the LREE of the system. By a redefinition of the 
temperature we observed that both entropies exactly are the
same. 

In comparison with the LREE of a  
dynamical-dressed stable D$p$-brane \cite{32},
presence of the tachyon field imposed 
the contributions of all mode numbers of the 
closed string to the LREE 
via an infinite product and an infinite series. 
Similar additional contributions also occurred 
for the thermodynamical entropy of the system.
As expected, when we turn off the tachyon field
the results of this paper completely reduce to 
the results of the dynamical-dressed stable
D$p$-brane \cite{32}.

We examined the behavior of the LREE 
under the condensation of the tachyon. 
We observed that the 
associated LREE to the D$p$-brane 
was decomposed to the 
LREE of a dynamical-dressed D$(p-1)$-brane and an  
entropy which may be corresponded  
to the released closed strings 
after the D$p$-brane collapse. The
latter entropy logarithmically
depends on the length of the lost direction
of the initial brane.  

Finally, because of the similar 
behavior of the thermal entropy and the LREE, we 
were persuaded to check the second 
law of thermodynamics for our LREE via 
the process of the tachyon condensation. 
We saw that preserving the 
second law obligates the parameters of 
the initial setup to obey 
a minimal condition. We explicitly
figured out the condition for the 
case of a D-string, which imposed 
a lower bound on the matrix element $U_{11}$.


\end{document}